# Earth's Alfvén wings driven by the April 2023 Coronal Mass Ejection

Li-Jen Chen[1], Daniel Gershman[1], Brandon Burkholder[1], Yuxi Chen[2], Menelaos Sarantos[1], Lan Jian[1], James Drake[3], Chuanfei Dong[2], Harsha Gurram[1,3], Jason Shuster[4], Daniel B. Graham[5], Olivier Le Contel[6], Steven J. Schwartz[7,8], Stephen Fuselier[9], Hadi Madanian[8], Craig Pollock[10], Haoming Liang[1,3], Matthew Argall[4], Richard E. Denton[11], Rachel Rice[1,3], Jason Beedle[4], Kevin Genestreti[4,9], Akhtar Ardakani[4], Adam Stanier[12], Ari Le[12], Jonathan Ng[1,3], Naoki Bessho[1,3], Megha Pandya[1,13], Frederick Wilder[14], Christine Gabrielse[15], Ian Cohen[16], Hanying Wei[17], Christopher T. Russell[17], Robert Ergun[8], Roy Torbert[4,9], James Burch[9]

1 NASA Goddard Space Flight Center, Greenbelt, MD, USA
2 Boston University, Boston, MA, USA
3 University of Maryland, College Park, MD, USA
4 University of New Hampshire, Durham, NH, USA
5 Swedish Institute of Space Physics, Uppsala, Sweden
6 CNRS/Ecole Polytechnique/Sorbonne Université/Univ. Paris Sud/Observatoire de Paris, Paris, France
7 Imperial College London, London, UK
8 Laboratory for Atmospheric and Space Physics, University of Colorado Boulder, Boulder, CO, USA
9 Southwest Research Institute, San Antonio, TX, USA
10 Denali Scientific, Healy, Alaska, USA
11 Dartmouth College, Hanover, NH
12 Los Alamos National Laboratory, Los Alamos, NM, USA
13 Department of Physics, The Catholic University of America, Washington, D.C., USA
14 University of Texas at Arlington, Arlington, TX, USA
15 The Aerospace Corporation, El Segundo, CA, USA
16 The Johns Hopkins University Applied Physics Laboratory, Laurel, MD, USA
17 University of California, Los Angeles, Los Angeles, CA, USA

**Plain Language Summary**
Like supersonically fast fighter jets creating sonic shocks in the air, planet Earth typically moves in the magnetized solar wind at super-Alfvénic speeds and generates a bow shock. Here we report unprecedented observations of Earth's magnetosphere interacting with a sub-Alfvénic solar wind brought by an erupted magnetic flux rope from the Sun, called a coronal mass ejection (CME). The terrestrial bow shock disappears, leaving the magnetosphere exposed directly to the cold CME plasma and the strong magnetic field from the Sun's corona. Our results show that the magnetosphere transforms from its typical windsock-like configuration to having wings that magnetically connect our planet to the Sun. The wings are highways for Earth's plasma to be lost to the Sun, and for the plasma from the foot points of the Sun's erupted flux rope to access Earth's ionosphere. Our work presents new opportunities to study interaction between astrophysical bodies with sub-Alfvénic plasma wind in our solar and other stellar systems.

**Abstract**
We report a rare regime of Earth's magnetosphere interaction with sub-Alfvénic solar wind in which the windsock-like magnetosphere transforms into one with Alfvén wings. In the magnetic cloud of a Coronal Mass Ejection (CME) on April 24, 2023, NASA's Magnetospheric Multiscale mission distinguishes the following features: (1) unshocked and accelerated cold CME plasma coming directly against Earth's dayside magnetosphere; (2) dynamical wing filaments representing new channels of magnetic connection between the magnetosphere and foot points of the Sun's erupted flux rope; (3) cold CME ions observed with energized counter-streaming electrons, evidence of CME plasma captured due to reconnection between magnetic-cloud and Alfvén-wing field lines. The reported measurements advance our knowledge of CME interaction with planetary magnetospheres, and open new opportunities to understand how sub-Alfvénic plasma flows impact astrophysical bodies such as Mercury, moons of Jupiter, and exoplanets close to their host stars.

**Introduction**
The Earth typically moves in the magnetized solar wind with super-Alfvénic speeds, generating a bow shock, magnetosheath, and a windsock-like magnetosphere with a tail. Occasionally and especially in the strong magnetic fields of the Sun's coronal mass ejections (CMEs) [Hajra and Tsurutani, 2022], the Alfvén speed exceeds the solar wind speed (sub-Alfvénic). MHD simulations [Ridley, 2007; Chané et al., 2015; Chen et al., 2024; Burkholder et al., 2024] have shown that during long duration (> 1 hour) sub-Alfvénic solar wind, Earth's magnetosphere transforms into Alfvén wings. Measurements from well-instrumented satellites orbiting Earth provide unique opportunities to study the physical processes of Alfvén-wing magnetospheres, such as those formed around Jupiter's moons [e.g., Kivelson et al., 2004] and extrasolar planets under sub-Alfvénic stellar wind [e.g., Pineda and Villadsen, 2023].

To date, observations of Earth's magnetosphere interaction with long-duration sub-Alfvénic solar wind have only been reported in three prior papers due to its rare occurrence. Magnetic field and ion flow measurements by Geotail at GSE [X~(-11: -20), Y~(28: 23), Z~5] $R_E$ captured the dusk Alfvén wing and an enhanced flow layer immediately outside of the reduced flow region of the wing [Chané et al., 2012]. The plasma properties (e.g., density, temperature, and energy distribution) of the enhanced flow layer were not measured. The observations by Lugaz et al. [2016] focus primarily on radiation-belt electron loss, and observed neither the enhanced flow layer nor the Alfvén wings. The work by Hajra and Tsurutani [2022] shows statistics of the sub-Alfvénic solar wind properties and associated changes in the geomagnetic fields in the inner magnetosphere as well the ionosphere, and does not address the Alfvén wings.

In this paper, we report dayside observations of Earth's Alfvén wing magnetosphere during the impact of a CME on April 24, 2023. Within the Sun's erupted magnetic flux rope of the CME is a low-density plasma parcel that brings the local Alfvén speed to above the solar wind speed for two hours, exposing the magnetosphere to sub-Alfvénic solar wind conditions. We use measurements from the Magnetospheric Multiscale (MMS) mission [Burch et al., 2016]. This event is the only time a long duration sub-Alfvénic solar wind occurs in the mission life of MMS so far. With its unprecedented high-cadence three-dimensional plasma measurements, MMS witnesses newly-generated Alfvén wing filaments interact as they connect the Sun's erupted flux rope with Earth's magnetic fields and facilitate solar-terrestrial plasma exchange.

**Observations**

For sub-Alfvenic solar wind with a dominant dawn-dusk (y) IMF, global MHD simulations [Chané et al., 2015; Chen et al., 2024] demonstrate the formation of dawn-dusk wings.
In the April 2023 CME at ~14 UT, MMS approaches the southern dawn sector on the dayside at GSE [10.7, -7.9, -6.8] $R_E$. The presented magnetic fields are measured by the Flux Gate Magnetometer [Russell et al., 2016], thermal plasma Fast Plasma Investigation [Pollock et al., 2016], and nonthermal particles Fly's Eye Energetic Particles spectrometers [Blake et al., 2016] onboard MMS2. Additional data from OMNI (King & Papitashvili, 2005), Wind spacecraft at Lagrange point 1 (Lepping et al., 1995), and Time History of Events and Macroscale Interactions during Substorms (THEMIS) spacecraft B (Angelopoulos, 2008) are presented to provide the context. Vectors are shown in the Geocentric Solar Ecliptic (GSE) coordinate system.

We begin with an overview of the CME in relation to the geomagnetic storm development (Figure 1). The fast-moving CME created an interplanetary (IP) shock and a sheath region in front of the expanding flux rope (Figure 1a), the magnetic cloud (MC). Arrival of the IP shock and the southward IMF afterward initiate the main phase of a geomagnetic storm, and as the IMF $B_z$ turns from southward to northward in the CME sheath (Figures 1b), the storm development stalls (Figure 1g). The CME sheath (bounded by the cyan and magenta vertical lines) is prominently delineated by (1) enhanced fluxes of 60-560 keV electrons and ions with the strongest fluxes and highest energy reached at the IP shock arrival (Figures 1c-1d), and (2) the IMF southward turning (Figure 1b) and density drop (Figure 1e). In the MC, MMS encountered Earth's bow shock five times (marked as red arrows in Figure 1b). MMS registers a double-peak density pulse at the second bow shock crossing (Figure 1e), coincident with a spike in the Alfvén Mach number $M_A = V_{sw}/V_A$ ($V_{sw}$ is the solar wind speed and $V_A$ is the Alfvén speed; Figure 1f), and a bipolar pulse in SYM-H (Figure 1g). This bow shock crossing has been analyzed in depth [Graham et al., 2024]. The density pulse consists of three ion species (a subject of forthcoming papers), and results in extreme responses of the geomagnetic field [Despirak et al., 2023; Zou et al., 2024].

MMS monitors the magnetosheath interaction with the MC continuously after its last bow shock crossing at ~4:20 UT (Figures 1b, 1e), and particularly as the solar wind $M_A$ drops and rises culminating in the two-hour interval (~12:30-14:30 UT) when $M_A$ stays at ~0.6 (Figure 1f). The sub-Alfvénic solar wind is caused by the low density and high IMF strength in the MC rather than a decrease in $V_{sw}$ (Figures 1e and 1h). The interval is concurrent with a SYM-H plateau at ~ -120 nT (Figure 1g), indicating that the ring current intensity is being held at a level stronger than that of most geomagnetic storms. We show the Wind magnetic field measurements (Figure 1a; without any time shift to preserve its original variations) to enable discernment of magnetic variations from the CME magnetic cloud and those due to interaction with Earth's magnetosphere. At approximately 14:30 UT, MMS encounters a region with the magnetic field components similar to those observed in the magnetosphere (~17 UT) and qualitatively different from the upstream fields measured by Wind and THEMIS B.

The magnetosheath plasma evolves from shocked (decelerated and heated; marked by the first magenta bar in Figure 2) to unshocked solar wind (beginning is marked with a blue arrow) just before 13 UT, and back to the shocked solar wind after ~1440 UT (second magenta bar). During the interval when $M_A<1$, the bow shock disappears [Ridley et al. 2007]. Characteristics of the unshocked solar wind include well-defined narrow energy bands of the cold proton and alpha particles (Figure 2c) at energies consistent with upstream measurements such as those from THEMIS B (at GSE [34.6, 48.5, 4.5] $R_E$ at 14 UT; Figure 2h). The unshocked solar wind flow registered by MMS shows an increase (Figure 2b), compared with the flat $V_{ixyz}$ profiles recorded by THEMIS B (Figure 2g), consistent with MHD simulation results [Chen et al., 2024]. The flow acceleration in the simulation is dominated by the JxB force, similar to the low-$M_A$ super-Alfvénic cases discussed by Lavraud and Borovsky [2008]. MMS measures the ions and electrons (Figure 2e) in the layer of enhanced flow to be cold and unthermalized low-density (Figure 2d), low-beta (~0.01) wind within the CME flux rope. This layer of accelerated cold CME plasma is now the immediate upstream of Earth's dayside open and closed field line regions as will be demonstrated in later figures. The enhanced flow layer has been demonstrated to be adjacent to the wings and magnetopause in MHD simulations [Chané et al., 2015; Chen et al., 2024], even though it was unknown whether the plasma in the layer could be heated.

The correlated discontinuities in magnetic field (Figure 2f) and ion energy distribution (2h) observed by THEMIS B orbiting the Moon correspond to Earth's bow shock crossing the spacecraft (right before and after 10UT, just after 11 UT, and after 16 UT). The bow shock expands beyond the Moon distance as $M_A$ approaches 1 from above.

In the zoom-in interval (Figures 2i-2o), MMS begins to encounter flux tubes with substantially more energetic plasmas than the unshocked CME MC plasma. Ions with energies ~20 keV are observed, while the MC protons are at ~3 keV and alpha particles at ~6 keV (Figure 2k). The flux of energized electrons peaks at 200-1000 eV, in contrast to the MC electron flux maximizing below 200 eV (for example ~14:07 UT; Figure 2m). The flux tubes correspond to excursions in the magnetic field (Figure 2i) and ion velocity (Figure 2j), energized or thermalized MC ions (Figure 2k), and density spikes (Figure 2l). These flux tubes are isolated (sandwiched by MC intervals) and detached from the magnetopause, based on multi-spacecraft timing and the following observation facts: (1) the flux tubes contain cold CME ions (Figures 2k, 3d-f, and 4f), distinct from magnetospheric ion populations, (2) the magnetic field inside the flux tube is more CME-like than magnetosphere-like (Figures 2a and 2i), (3) ion flows inside the flux tubes tend to have larger $V_y$ (Figure 2j) than the ambient MC flow along -y, not consistent with MMS going in and out of the magnetopause. We name the isolated flux tubes (14:05-14:24 UT) filaments. The filaments contain field lines with at least one end connected to the Earth, an interpretation based on the assumption that the energized electrons and ions are either of Earth origin or due to interaction between the MC and the magnetosphere such as reconnection. We will further zoom in example filaments in Figure 4.

During the sub-Alfvénic interval, the MC magnetic field is northward and dominated by a dawnward $B_y$ (Figure 2a). $B_{xyz}$ at ~ 14:25-35 UT are distinct from those in the MC (Figure 2i) and similar to those in the magnetosphere at ~17 UT. Moreover, this interval displays that (1) ion flows are stagnant or weak (Figure 2j), and ions and electrons are hot (Figures 2k and 2m), (2) the density (0.25 cm$^{-3}$) is lower than the MC density (0.5 cm$^{-3}$), (3) enhanced fluxes of 60-150 keV electrons (Figure 2n) peak at pitch angles approximately 90 degrees (Figure 2o). The above properties are consistent with the interval containing primarily closed field lines. This interpretation is further supported by approximately balanced counter-streaming keV electron fluxes (part of the interval will be shown in Figures 3b1-c1).

Prior to and after the interval shown in Figures 2i-2o, MMS registers the MC plasma and fields. Upon exit from the magnetosphere and back into the enhanced MC flow layer, MMS detects the MC proton energy at 3-4 keV, corresponding to velocities 750-850 km/s, and then drops to 1-2 keV (Figure 2k). The ion energy drop occurs with a sudden density increase (Figure 2l), signaling the end of the sub-Alfvénic interval locally at MMS. After the density rise, ions regain their "shocked" appearance. A corresponding density rise is observed by Wind and THEMIS B spacecraft (not shown) at L1 and the Moon, respectively.

At entry (first green bar in Figure 2l) into and exit (second green bar) from the closed field line region, MMS observes features consistent with Alfvén-wing and freshly-closed field lines. In Figure 3 during the MC intervals (marked with orange arrows), electrons are counter-streaming, and the anti-parallel flux due to the strahl component extends to several hundred eV, higher than the parallel electron energies. The presence of strahl electrons indicates magnetic connection to the solar corona [e.g., Borovsky, 2021]. In the red interval, the anti-parallel electron flux is MC-like, while the parallel flux shows higher energy electrons in addition to the MC-like population (Figures 3b1-c1), consistent with dawn wing flux tubes generated by IMF reconnecting with closed field lines southward/duskward of MMS. The green interval and thereafter (until ~14:35 UT) show counter-streaming electrons extending above 1 keV with no discernable MC-like populations, interpreted as closed field lines. The ions, although decelerated and heated to various degrees, exhibit a trace reminiscent of the MC ion population (most visible in Figure 3f1), while the corresponding magnetic field (Figure 2i) and ion flow (Figure 2j) are transitioning from MC-like to magnetosphere-like. Inspired by previous studies on magnetopause boundary layers under northward IMF and super-Alfvénic solar wind [e.g., Song and Russell, 1992; Lavraud et al., 2006; Fuselier et al., 2014], we interpret intervals with roughly-balanced bi-directional keV electrons as flux tubes recently (during the sub-Alfvénic interval to account for traces of MC ions) closed due to IMF reconnection with the dusk- and dawn-wing field lines, termed dual-wing reconnection here. Dual-wing reconnected flux tubes of various ages correspond to cold MC ions being decelerated and heated to various degrees.

Significant differences in the energy distribution and/or flux intensity of the bi-directional keV electrons (such as the interval right after the blue arrow) could be due to different reconnection strengths on two ends of the flux tube, or closed field lines being re-opened on one end by new reconnection. Examination of the 3D electron distribution functions is required to conclusively differentiate the two scenarios, and will be the subject of future investigations.

At exit back to the MC, cold CME ions prevail through much the shown interval. The blue interval has cold MC-like ions (Figures 3a2 and 3d2-f2), MC-like parallel electron energy distribution (Figure 3b2), and anti-parallel keV electrons superposed on a slightly depleted MC electron population (Figure 3c2), consistent with a newly formed dusk wing flux tube due to reconnection (IMF with closed field lines) northward of MMS. The green interval contains energized bi-directional electrons in addition to partially depleted MC electrons, consistent with dual-wing reconnection capturing MC plasmas on closed field lines. Near the end of the interval at ~14:38:15 UT are cold MC ions and counter-streaming MC electrons with the anti-parallel strahl component.

In Figure 4, we zoom in example filaments in the interval 14:08:00 to 14:09:20 UT (red bar in Figure 2l) to show their structures and dynamics. In the MC intervals (e.g., immediately before the blue and after the red intervals), the cold proton population is at ~2 keV (Figure 4b), and electrons (Figures 4c-d) display the same counter-streaming and strahl signatures as in Figure 3. The filament magnetic fields remain largely MC-like (Figure 4a), and the flows within have stronger dawnward component than the MC flow. All filaments exhibit flux enhancements of energized keV electrons (Figure 4c-d). The electron density varies by up to a factor of two (Figure 4e). The ion phase space distribution in $v_{ix}$ (Figure 4f) indicate that the cold CME ion population is discernable throughout the filaments despite being decelerated, accelerated, and even heated to a certain degree. We highlight one example interval for each distinct magnetic topology. The blue interval shows cold MC-like ions, MC-like parallel electrons, keV anti-parallel electrons superposed on a mostly depleted MC electron population, consistent with a newly formed dusk-wing flux tube due to IMF reconnecting with closed field lines northward of MMS. The green interval contains cold as well as slightly heated MC ions (Figure 4b and 4f), and bi-directional keV electrons on MC-like magnetic fields, interpreted as dual-wing reconnected. The red interval displays MC-like anti-parallel electrons, substantially more energetic parallel electrons, and cold CME ions, interpreted as a dawn-wing flux tube newly generated by IMF reconnecting with closed field lines duskward/southward of MMS.

To facilitate visualization, we illustrate field lines with distinct topologies in Figure 4h based on data from a global MHD simulation [Chen et al., 2024]. The field lines are traced from the surface of r = 3 $R_E$ from uniformly distributed seed points covering the entire 3D sphere, representative of all possible field line topologies connected to the magnetosphere in the simulation. The open field lines extending from the southern hemisphere towards dawn (red) and northern hemisphere towards dusk (blue) form two wings connecting the Earth and the Sun magnetically. The most probable sites of IMF reconnecting with magnetosphere field lines (green) are southern dusk and northern dawn in the simulation. Example sites are identified by pairs of newborn dawn- and dusk-wing field lines (indicated by yellow arrows at southern dusk and white field lines with sharp kinks at northern dawn). The southern dusk reconnection site is of order 10 $R_E$ away from MMS. For reference, electrons at 1 keV travel this distance in 1.8 s.

With the observed magnetic topologies in mind, we consider the implied physical picture. The dawn-wing (dusk-wing) intervals with MC electrons in the anti-parallel (parallel) direction present evidence for MMS being on field lines connected to the solar corona. The freshly closed field lines are distinguished by their MC-like magnetic field (not yet oriented to the expected magnetosphere field configuration), MC-like cold ions, and sometimes even MC-like electron distributions superposed on the keV populations (e.g., green interval in Figures 3a2-f2). Part of wing field lines in a flux tube may reconnect with field lines from the other wing, leading to a single filament with multiple topologies. Note that multiple flux tubes of one topology often mingle with those of other topologies, suggesting that the flux tubes of distinct topologies move with respect to one another, merge or reconnect, and are being assimilated into the magnetosphere and wings.

**Summary and Discussion**

In this paper, we report a rare regime of the terrestrial magnetosphere interaction with sub-Alfvénic CME plasma. The immediate magnetosphere upstream is cold and accelerated CME plasma on magnetic-cloud field lines, instead of the usual magnetosheath with heated and decelerated solar wind. The beta in the CME magnetic cloud reaches as low as 0.01, in contrast to the typical magnetosheath beta > 1. Lower beta has been demonstrated to support faster reconnection [Zenitani and Miyoshi, 2020], stronger plasma energization [Phan et al., 2013a; Li et al., 2021], and higher reconnection occurrence rates [Phan et al., 2010, 2013]. The magnetopause reconnection with the low beta CME solar wind as its upstream resembles that at the inner planets (such as Mercury [e.g., Sarantos and Slavin, 2009; Lavraud and Borovsky, 2008]) of the solar system and at magnetized exoplanets under sub-Alfvénic stellar wind [Saur et al., 2013; Pineda and Villadsen, 2023].

The sub-Alfvénic nature of the CME solar wind allows newly reconnected flux tubes to move as isolated filaments detached from the magnetopause, increasing interaction opportunities between field lines of distinct magnetic topologies, and enabling efficient reconnection. Cold CME plasmas are captured on closed field lines, likely by dual-wing reconnection. Dual-wing reconnection representing draped IMF reconnecting with dusk and dawn wing field lines at northern-dawn and southern-dusk cusps, respectively, is analogous to dual-lobe reconnection for super-Alfvénic solar wind in northward IMF [Song and Russell, 1992; Lavraud et al., 2006; Hasegawa, 2012; Fuselier et al., 2014]. As sub-Alfvénic wind does not force newly reconnected flux tubes to convect tailward, various reconnection scenarios, including dual-wing reconnection at locations other than cusps and reconnection between wing field lines as well as with newly closed field lines of different ages, become possible. The highly dynamic process of flux tubes with varying topologies and ages forming, moving, and reconnecting with one another accounts for MMS observations of multiple magnetic topologies co-existing in filaments. Under super-Alfvénic solar wind conditions, lobe field lines convect tailward and cannot meet on the dayside, a key distinction from dual-wing reconnection.

The data reported in this paper indicate that the geo-effectiveness of sub-Alfvénic solar wind remains an open question. SYM-H stays approximately constant at -120 nT for the sub-Alfvénic interval, prolonging the geomagnetic storm recovery and suggesting that injections of energetic ions may occur to replenish the ring current. These injections are possible consequences of near-Earth nightside magnetic reconnection predicted for the Alfvén-wing magnetosphere [Burkholder et al., 2024; Chen et al., 2024].

Sun-Earth connection through Alfvén wings is analogous to the Jupiter-Ganymede connection. MMS measures newly generated Alfvén wing flux tubes, representing channels of magnetic connection between Earth's magnetosphere and the foot points of the Sun's erupted flux rope. MMS observations indicate that the Earth loses its plasma to the solar corona through the Alfvén wings established by magnetic reconnection, and suggest that the Ganymede aurora on Jupiter is likely powered by reconnection as well. Accelerated field-aligned electrons at the magnetopause and auroral emissions at Ganymede support reconnection occurrence [Ebert et al, 2022; Gershman et al., 2024]. Conversely, electrons from the foot points of the Sun's erupted flux rope may be accelerated and precipitate into Earth's ionosphere to form Alfvén wing aurora – another form of geo-effectiveness to be checked out by future work.

**Open Research**
MMS data are publicly available at https://lasp.colorado.edu/mms/sdc/public/about/browse-wrapper/. Wind and THEMIS B data are obtained from the CDAWeb https://cdaweb.gsfc.nasa.gov/index.html/.

**Acknowledgments**
This work is supported by the NASA MMS Mission. LJC thanks Eric Grime for his guidance on pyspedas, and Hiroshi Hasegawa for discussions on reconnection scenarios to interpret electron data.

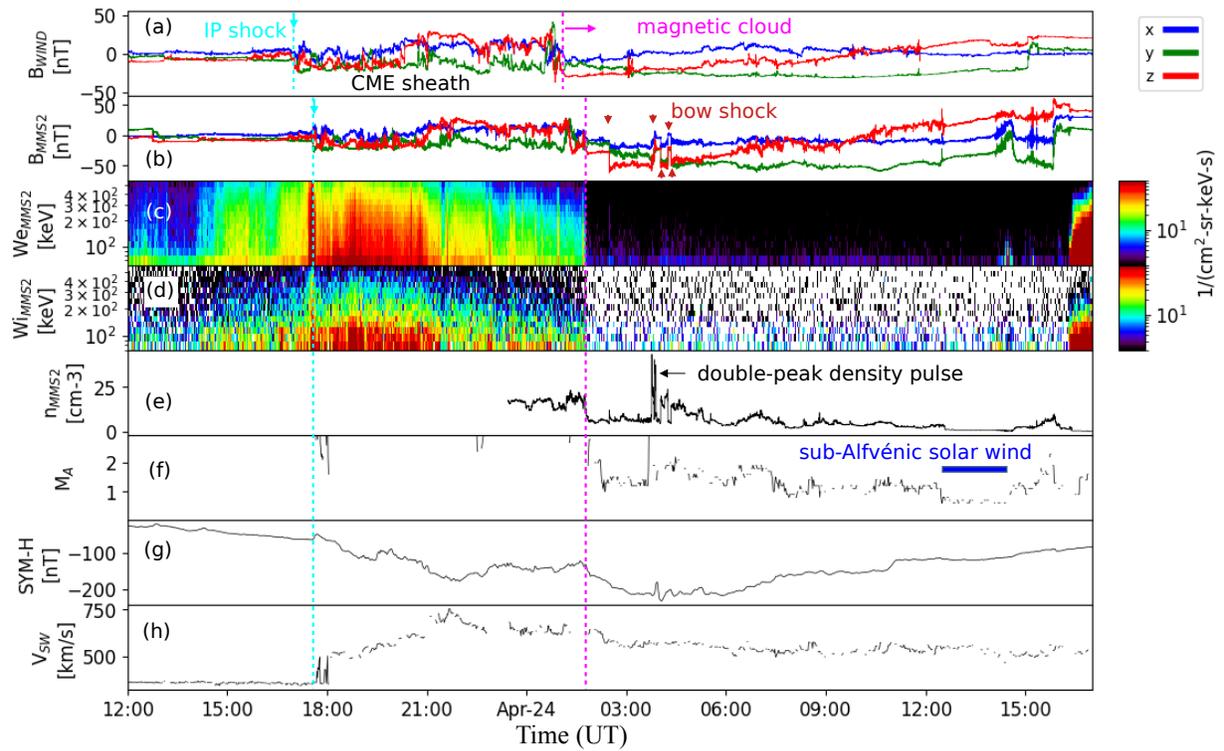

Figure 1. Overview of the 2023 April CME and the associated geomagnetic storm. (a) Three components of the magnetic field observed by the Wind spacecraft at L1, and (b) by MMS2 on the dayside, respectively. (c-d) MMS suprathermal electron and ion (60-560 keV) energy fluxes. (e) MMS electron density. (f) Alfvén Mach number $M_A$ (from the OMNI data which have been propagated to the model bow shock nose). (g) Geomagnetic storm index SYM-H. (h) Solar wind speed from OMNI.

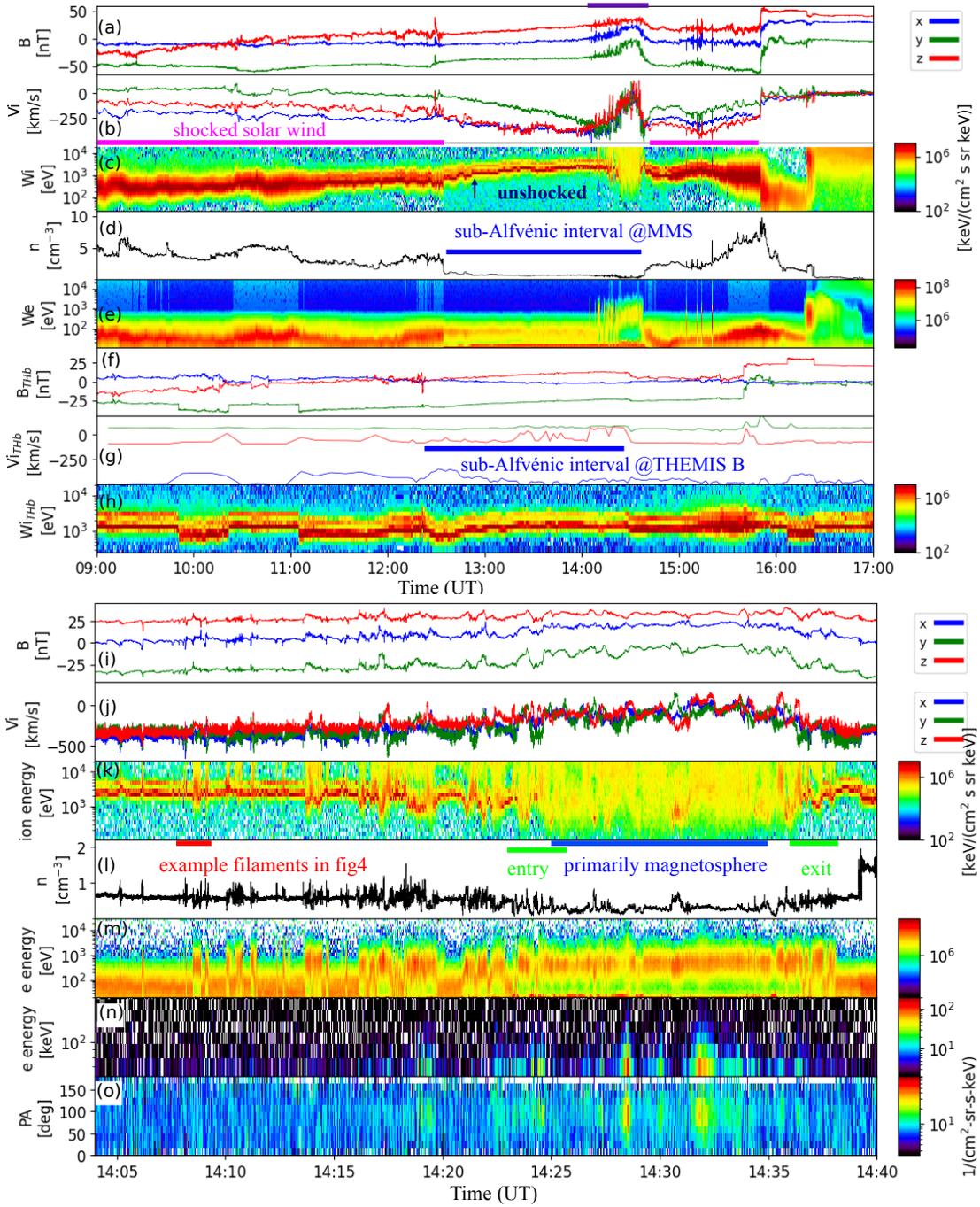

Figure 2. MMS observation of the magnetosheath evolution, Alfvén wing filaments, and storm-time magnetosphere. (a) Magnetic field. (b) Ion velocity. (c) Ion energy flux. (d) Electron density. (e) Electron energy flux. (f-h) Magnetic field, ion velocity, and energy flux from THEMIS B (orbiting the Moon at GSE [34.6, 48.5, 4.5] $R_E$ at 14 UT) to provide a near-Earth upstream context. The MMS interval 14:04-14:40 UT (purple bar) is zoomed in to highlight the filaments and distinct electron regions, based on (i) magnetic field, (j) ion flow, (k) energy flux of ions (0.2-20 keV), (l) plasma density, (m) energy flux of electrons (0.02-3 keV), and (o-p) energetic electron flux (60-200 keV) and their pitch angle (PA) distribution.

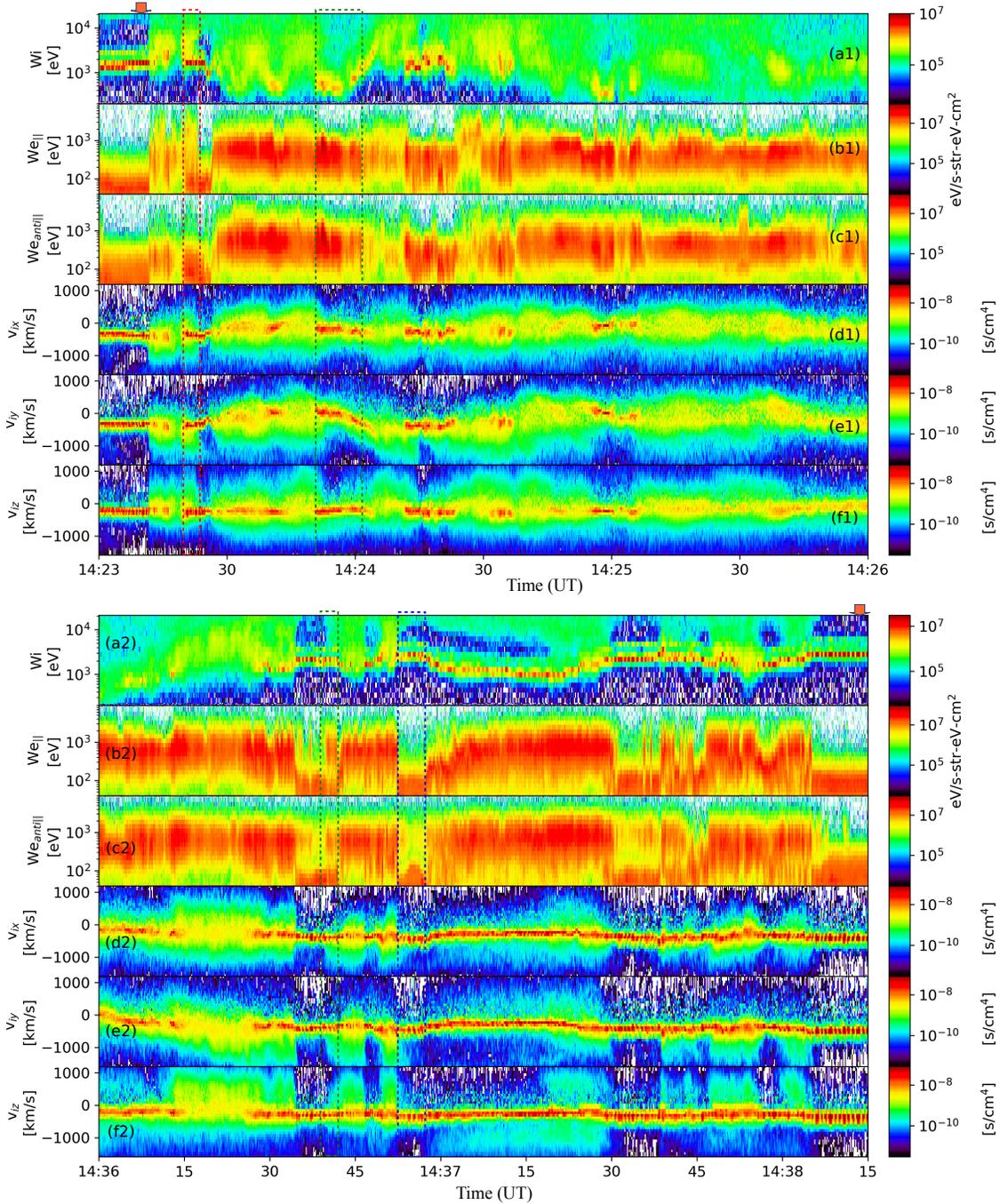

Figure 3. Evidence for the Alfvén-wing flux tubes and freshly closed field lines. Entry (a1-f1) into and exit (a2-f2) from the closed field line region of the magnetosphere. (a1,a2) ion energy flux to provide a context for distinct plasma regions. (b1,b2) electron energy flux parallel to **B**. (c1,c2) electron energy flux anti-parallel to **B**. Ion phase space density (summed over the other two velocity dimensions) as a function of $v_{ix}$ (d1,d2), $v_{iy}$ (e1,e2), and $v_{iz}$ (f1,f2). The red (blue) interval has keV electrons parallel (anti-parallel) to **B**, and anti-parallel (parallel) MC electrons, representing the dawn (dusk) wing flux tubes. The green intervals have cold CME ions with bi-directional keV electrons, consist with flux tubes generated by dual-wing reconnection. Orange arrows mark example time with MC electrons.

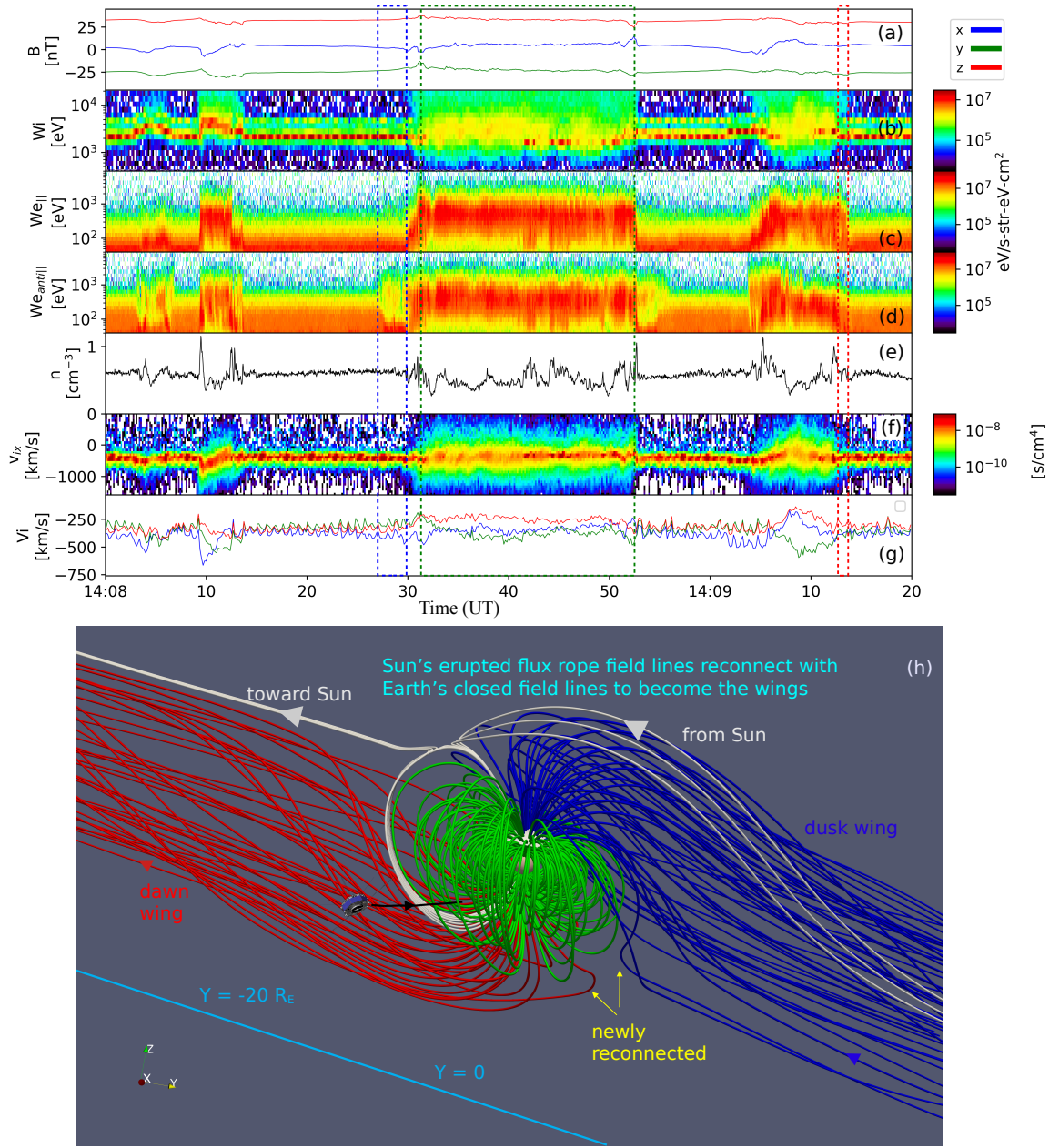

Figure 4. Zoom-in of example filaments and a global field-line illustration. (a) Magnetic field. (b) Energy flux of ions (0.4-20 keV) summed over all directions. (c-d) Energy flux of electrons (0.02-30 keV) parallel and antiparallel to **B**. (e) Electron density. (f) Ion phase space density (summed over $v_{iy}$ and $v_{iz}$) as a function of $v_{ix}$. (g) Ion flow vector. (h) Illustration of the CME flux rope field lines interacting with Earth's Alfvén-wing magnetosphere based on an MHD simulation. The southern-dawn wing is in red, northern-dusk wing blue, closed field lines green, and example newly reconnected as well as soon-to-be-reconnected field lines white. The black line segment with an arrow represents the approximate trajectory of MMS. The quasi-periodic oscillations in the ion data (panels b, f, and g) are instrumental.